\documentclass{emulateapj}
\newcommand\etal{{\it et al.~}}
\newcommand\beq{\begin{equation}}
\newcommand\eeq{\end{equation}}
\newcommand\beqar{\begin{eqnarray}}
\newcommand\eeqar{\end{eqnarray}}

\newcommand\bvec[1]{\hbox{\boldmath${#1}$}}

\newcommand{\cref}{C_{\rm ref}}

\begin{document}

\title{PROTOSTAR FORMATION IN MAGNETIC MOLECULAR CLOUDS BEYOND ION DETACHMENT:
\\II. TYPICAL AXISYMMETRIC SOLUTION}

\author{Konstantinos Tassis\altaffilmark{1,2} \& Telemachos Ch. Mouschovias\altaffilmark{1}}

\altaffiltext{1}{Departments of Physics and Astronomy,
University of Illinois at Urbana-Champaign, 1002 W. Green Street, Urbana, IL 61801}
\altaffiltext{2}{Department of Astronomy and Astrophysics and the 
Kavli Institute for Cosmological Physics, 
The University of Chicago, 5640 South Ellis Avenue,
Chicago, IL 60637}

\begin{abstract}

We follow the ambipolar-diffusion--driven formation
and evolution of a fragment in a magnetically supported
molecular cloud, until a hydrostatic protostellar core forms at its center. This problem was formulated in Paper I. We determine the density, velocity and magnetic field as functions of space and time, and the contribution of ambipolar diffusion and Ohmic dissipation to the resolution of the magnetic flux problem of star formation. The issue of whether the magnetic field ever decouples from the (neutral) matter is also addressed. We also find that the electrons do not decouple from the field lines before thermal ionization becomes important and recouples the magnetic field to the neutral matter. Ohmic dissipation becomes more effective than ambipolar diffusion as a flux reduction mechanism only at the highest densities (a few $\times 10^{12}$ ${\rm cm^{-3}}$). In the high-density central parts of the core, the magnetic field acquires an almost spatially uniform structure, with a value that, at the end of the calculation ($n_{n} \approx 5 \times 10^{14}$ ${\rm cm^{-3}}$), is found to be in excellent agreement with meteoritic measurements of magnetic fields in the protosolar nebula. Outside the hydrostatic protostellar core, a {\em concentration of magnetic flux} (a ``magnetic wall'') forms, which gives rise to a magnetic shock. This magnetic shock is the precursor of the repeated shocks found by Tassis \& Mouschovias (2005b) which cause spasmodic accretion onto the hydrostatic core at later times.

\end{abstract}


\section{Introduction}

In a companion paper (Tassis \& Mouschovias 2006a, hereafter Paper I) we formulated the problem of 
the formation, by ambipolar diffusion, and evolution of a magnetically supercritical molecular
 cloud fragment, past the phase of ion detachment from the magnetic field lines (leaving the electrons 
as the only attached species). The formulation can follow the evolution of the core in the opaque regime 
until the central temperature rises to about 1000 K. At higher temperatures, thermal ionization becomes 
important and the chemical model no longer accurately predicts the degree of ionization in the 
core\footnote{In addition, new physics is expected to become important soon after  
this point; e.g., significant deviation from disklike geometry in the central region; formation of 
a centrifugal barrier and an associated rotationally supported protostellar disk.}. However, by the 
time thermal ionization becomes important and the increase of the degree of ionization reattaches the 
magnetic field to the matter, the question of whether the magnetic field fully detaches 
from the matter at any point in the evolution of a forming protostar will already have been answered. 

During the opaque phase of evolution, several important effects are expected to set in. First, if 
there is ever going to be a complete decoupling of the magnetic field from the matter, it must occur 
after the detachment of the ions but before thermal ionization becomes important. (Complete decoupling 
refers to conditions such that the magnetic field has no significant effect on the dynamics of the 
neutral matter, and the motion of the neutral matter no longer affects the magnetic field.) Hence, detachment of the last 
attached species (the electrons) from the magnetic field is necessary for magnetic decoupling to 
occur. Second, it is likely that a significant part of the magnetic-flux reduction necessary for 
the resolution of the magnetic flux problem of star formation will occur in this phase, due to 
ambipolar diffusion or Ohmic dissipation, or both. (Desch \& Mouschovias 2001 showed that ambipolar 
diffusion is resurrected during those late stages of contraction of the supercritical core.) Third, 
the formation of the ``magnetic wall'', found by \cite{TassisM05a,TassisM05b} to be responsible 
for the spasmodic accretion onto the formed protostellar core, should manifest itself at this 
advanced stage of evolution. 

In this paper, we investigate these effects as well as the structure of the evolving core for 
typical values of the free parameters, described below. A detailed parameter study is presented 
in the following paper (Tassis \& Mouschovias 2006c, hereafter Paper III). In \S \, \ref{pm} 
we summarize the physical processes and assumptions relevant to the model cloud, and describe 
the free parameters of the model and their values. In \S \, \ref{mainrun} we describe the results 
for the structure and evolution of the core, the late-time behavior of the magnetic field, and we 
discuss the resolution of the magnetic flux problem of star formation. The main conclusions are 
summarized and discussed briefly in \S \, \ref{disc}.

\section{The Physical Model}\label{pm}

The details of the physical processes included in the calculations were
presented in Paper I. In summary, a model cloud is an
axially symmetric, nonrotating, thin disk embedded in an external medium of constant 
thermal pressure $P_{\rm ext}$, with the axis of symmetry aligned with the $z-$axis of a
cylindrical-polar coordinate system $r,\phi,z$. 
We take advantage of previous calculations \citep{FM92, FM93, DM01}, which show that a
typical model cloud relaxes rapidly along magnetic field lines and achieves
balance between thermal-pressure and gravitational forces in this direction. 
This balance of forces is maintained throughout the evolution. Hence, the dimensionality of
the MHD equations can be reduced by integrating analytically over $z$, allowing for
a much more efficient numerical solution, without loss of any physics
relevant to the problem.

The magnetic flux tubes are initially loaded with mass by specifying a reference state
which has a uniform magnetic field $B_{\rm ref}$ and a column density 
\beq \label{Tpro}
\sigma_{\rm n,ref}=\frac{\sigma_{\rm c, ref}}
{\left[1+(r/l_{\rm ref})^2\right]^{3/2}}\,,
\eeq
where $\sigma_{\rm c, ref}$ and $l_{\rm ref}$ are the central column
density and radial lengthscale, respectively \citep{Toomre63}. 
The cloud boundary is placed initially at $r = R_0 = 5 l_{\rm ref}$, where the column
density falls to less than $1\%$ of its central value. The reference
state is isothermal, at a temperature of $10$ K, corresponding to an
isothermal sound speed $C=0.188 {\rm \,\, km \,s^{-1}}$.

The system of equations ([22a-n] in Paper I) governing the evolution of the model cloud
contains the following (dimensionless) free parameters:
 $\tilde{P}_{\rm ext}$,  $\tilde{l}_{\rm ref}$, $\mu_{\rm d,c0}$.
The parameter  $\tilde{P}_{\rm ext}$, which is the ratio of the
external pressure and the ``gravitational pressure'' ($=\pi G \sigma_{\rm c,ref}^2/2$) 
of matter in the central flux tube of the reference state,
enters the calculation through the condition of thermal-pressure balance across the cloud surface. The typical value used in this run is $\tilde{P}_{\rm
  ext}=0.1$. The parameter  $\tilde{l}_{\rm ref}$, which is the
dimensionless lengthscale introduced by the profile of the column density in the reference state (eq. [\ref{Tpro}]), is a measure of the strength of the initial
thermal-pressure force relative to the gravitational force inside the cloud. The value used in this run is $\tilde{l}_{\rm ref} = 5.5 \pi > 3\sqrt{2}$, which makes the cloud significantly thermally supercritical. The parameter  $\mu_{\rm d,c0}$,
which is the initial central mass-to-flux ratio in units of the
critical value for collapse, has a value $\mu_{\rm d,c0}=0.25$,
which makes the model cloud magnetically subcritical. 

The above values of the free parameters can be obtained from the following set of initial values of the physical quantities characterizing the reference state: $n_{\rm n,c0} = 2600 {\rm \, \, cm^{-3}}$; $B_{\rm ref} = 35.26 {\rm \, \, \mu G}$; $l_{\rm ref} = 0.858 {\rm \, \, pc}$; $R_0 = 4.3 {\rm \,\, pc}$; $M_{\rm tot} = 99 {\rm \,\,M_\odot}$. Parameters stemming from the microscopic processes (ionization,
chemistry, collisions) are very well constrained observationally and
experimentally, and typical values used in the calculations presented
here are given in Appendix A of \cite{TassisM05a}.

An initial equilibrium state is calculated by allowing the cloud to
relax under the assumption of flux-freezing. However, because the
cloud is initially magnetically subcritical, only  mild readjustments
of the column density and magnetic field profiles of the reference state 
take place before the equilibrium state is reached.

The opaque phases of the evolution are treated using the adiabatic
approximation above a certain density $n_{\rm opq}$, which is an
additional free parameter in the model. In the case of the
typical model cloud, $n_{\rm opq} = 10^{11} {\,\,\rm cm^{-3}}$.
Detailed studies of the effect of different values of the free
parameters on the evolution of the model cloud are presented in Paper III. 

The use of an adiabatic equation of state is 
a limiting case, representing the maximum rate at which the temperature 
can increase due to gravitational contraction\footnote{In principle, ambipolar diffusion and Ohmic dissipation may also contribute to the heating of the fluid. 
Then the maximum rate at which the temperature could increase is determined by the rate of compressional heating (due to gravitational contraction) {\em plus} the heating rate due to magnetic-diffusion processes. Then we find on the basis of the simulation described in this paper that heating due to ambipolar diffusion and Ohmic dissipation is five orders of magnitude smaller than compressional heating.} 
- no radiative losses are 
allowed. The other extreme, in which radiative losses are assumed 
to be efficient enough for isothermality to be maintained throughout 
the calculation, is presented in Paper III. The results of an exact calculation, 
in which the evolution of the temperature is modeled through detailed 
radiative transfer calculations, will necessarily be bracketed 
by the results of these two extreme cases.

\begin{figure*}
\plotone{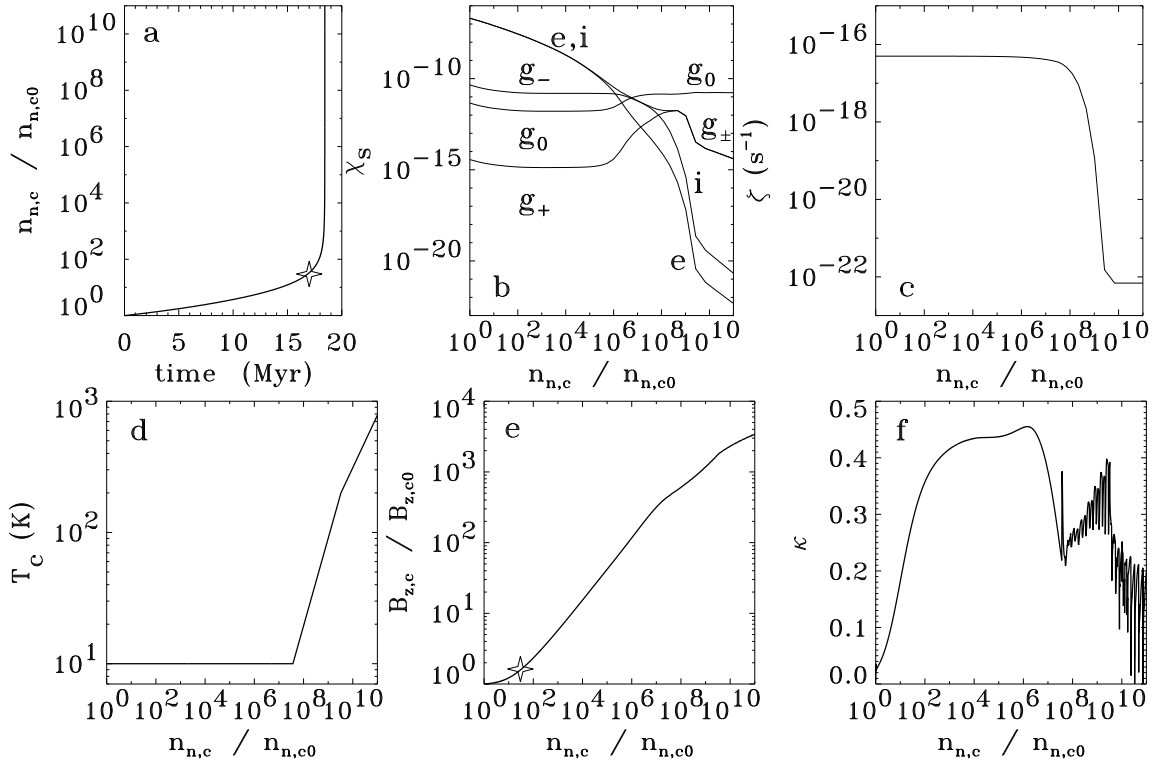}
\caption{\label{typ_cen_1} 
(a) Central number density of neutrals, $n_{\rm n,c}$, normalized to the
central number density of the initial equilibrium state 
($n_{\rm n,c0} = 2711$ ${\rm cm^{-3}}$),  as a function of time. 
The ``star'' marks the time of formation of the supercritical core.
 Evolution of central quantities of the typical model cloud as functions of the central neutral density enhancement, $n_{\rm n,c}/n_{\rm n,c0}$, where $n_{\rm n,c0} = 2711$ ${\rm cm^{-3}}$:
(b) Abundances of species. 
(c) Ionization rate at the center of the cloud due to all
processes (cosmic rays and radioactive decays).  
(d) Central temperature (in K). 
(e) $z-$component of the magnetic field, normalized to its value in the initial
equilibrium state ($B_{z,{\rm c0}}=36$ $\mu G$).
(f) Exponent $\kappa$ in the relation $B_{z, {\rm c}}\propto n_{\rm n,c}^\kappa$.
}
\end{figure*}

\subsection{Flux-Reduction Timescales}

During the late stages of the contraction of a magnetically supercritical core, 
the mass-to-flux ratio in the central flux tubes 
changes with time, as a result of the combined effects of ambipolar diffusion and Ohmic 
dissipation. In order to gauge the relative importance of these two flux-reduction mechanisms
and to evaluate their contribution to the resolution of the magnetic flux problem of star 
formation, it is useful to define characteristic timescales for the two processes, that can be calculated at any timestep from the values of physical quantities and their spatial derivatives.

In the geometry assumed in the present work, the overall timescale in which the local mass-to-flux 
ratio evolves, $\tau_{\rm \Phi_{\rm B}}$, is defined by
\beq\label{eqdef}
\tau_{\rm \Phi_{\rm B}} = \frac{\Phi_{B}}{\left | d \Phi_{B}/dt \right |}\, ,
\eeq
where $d\Phi_B/dt = \partial \Phi_B/\partial t + \bvec{v}_{\rm n} \cdot \nabla \Phi_B$ is the time derivative comoving with the neutrals. Following the procedure in \cite{m91} (eq. [30]), we add and subtract $\bvec{v}_{\rm f} \cdot \nabla \Phi_B$ and we use the fact that $\bvec{v}_{\rm f}$ 
(the effective flux advection velocity, including any flux redistribution due to Ohmic dissipation, 
see paper I)
defines a frame in which $\Phi_B$ is conserved, to find that
\beq
\frac{d\Phi_B}{dt} = (\bvec{v}_{\rm n} - \bvec{v}_{\rm f}) \cdot \nabla\Phi_B = 
(v_{\rm n} - v_{\rm f})\frac{\partial \Phi_B}{\partial r}\,,
\eeq
where the last step assumes axial symmetry and no drift velocities between species inside the disk along field lines (i.e., the $z$-direction). Using the generalized Ohm's law (eq. [17], Paper I) $\bvec{v}_{\rm f}$ can be expressed as
\beq
v_{\rm f} = v_{\rm p} + \frac{c (j_{\phi}+j_{0,\phi})}{B_z \sigma_{\rm p}}\,, 
\eeq
where $v_{\rm p}$ is the flux advection velocity and $\sigma_p$ the plasma conductivity
(see paper I for a detailed discussion of their definitions). 
Hence, equation (\ref{eqdef}) yields
\beq\label{mtftwoterms}
\frac{1}{\tau_{\Phi_{\rm B}}} = \frac{1}{\Phi_B} (v_{{\rm p}r}-v_{{\rm n}r}) \frac{\partial \Phi_{B}}{\partial r} + \frac{1}{\Phi_B} \frac{c}{B \sigma_{\rm p}} 
(j_{\phi} + j_{0 \phi}) \frac{\partial \Phi_{B}}{\partial r} \, .
\eeq

The contributions of ambipolar diffusion and Ohmic dissipation to the overall 
flux reduction represented in the above equations can be separated on the basis of a
simple geometrical argument. The term ``Ohmic dissipation'' refers to reduction of the magnetic flux due to disruption of the {\em current}. In the assumed geometry, currents can only exist in the $\phi-$direction ($j_z$ and $j_r$ vanish identically). Ambipolar diffusion, on the other hand, refers to a redistribution of magnetic flux due to a drift between the field lines (and any charged species attached to them) and the neutral particles. However, in this particular geometry, only drifts in the $r-$direction 
result to a redistribution of mass among different flux tubes. Drifts in the $\phi-$direction can only shift mass within the same flux tube, and do not alter the mass-to-flux ratio of the particular flux tube.

So, the first term on the right-hand side of equation (\ref{mtftwoterms}) is the
inverse of the timescale of the mass-to-flux ratio change due to
ambipolar diffusion, while the second term is the
inverse of the timescale of the mass-to-flux ratio change due to Ohmic
dissipation. Hence, we may write 
\beq
\frac{1}{\tau_{\rm AD}} =
\frac{1}{\Phi_B} (v_{{\rm p}r}-v_{{\rm n}r}) \frac{\partial \Phi_{B}}{\partial r}
\eeq
and 
\beq
\frac{1}{\tau_{\rm OD}} = 
\frac{1}{\Phi_B} \frac{c}{B \sigma_{\rm p}} 
(j_{\phi} + j_{0 \phi}) \frac{\partial \Phi_{B}}{\partial r} \, ,
\eeq
so that
\beq
\frac{1}{\tau_{\rm \Phi_{\rm B}}}=\frac{1}{\tau_{\rm AD}}+ \frac{1}{\tau_{\rm OD}}\,,
\eeq 
as expected for processes occurring in parallel \citep{mp86}.

\section{Results}
\label{mainrun}

\begin{figure*}
\plotone{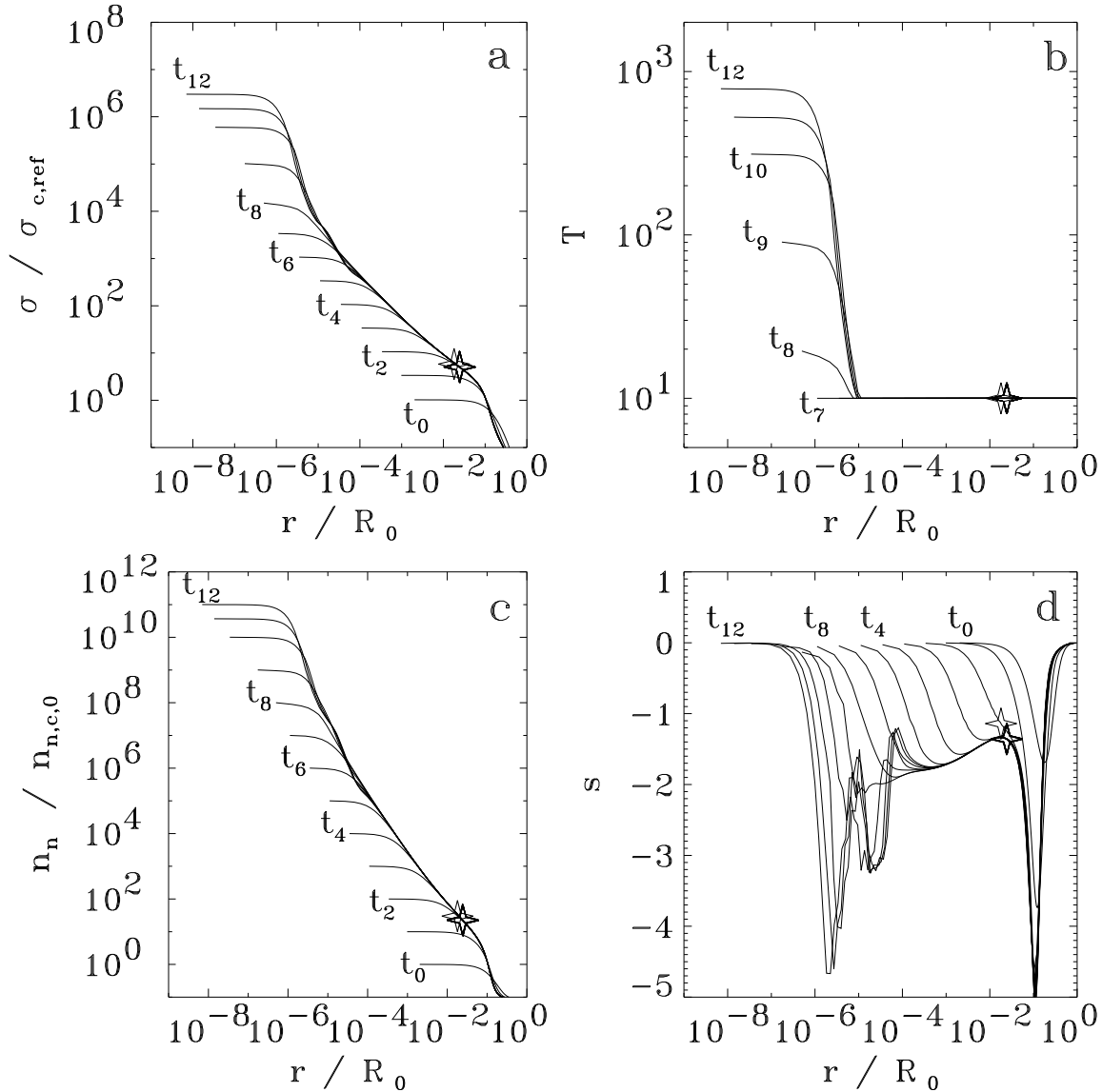}
\caption{\label{typ_rad_rho} 
Radial profiles of physical quantities at thirteen different times, ($t_0 = 0$,  $t_1 = 14.37$, $t_2 = 17.9$, $t_3 = 18.35$, $t_4 = 18.42$, $t_5 = 18.432$, $t_6 = 18.434$, 
$t_7 = 18.435$, $t_8 = 18.4354$, $t_9 = 18.4358$, $t_{10} = 18.4362$, $t_{11} = 18.4364$, $t_{12} = 18.4366$).
(a) Column density, normalized to the central column density of the reference
state, $\sigma_{\rm c, ref} = 5.58\times10^{-3}$ ${\rm g \, cm^{-2}}$. 
(b) Temperature in ${\rm K}$.
(c) Central number density of neutrals, $n_{\rm n,c}$, normalized to the
central number density of the initial equilibrium state, as in Fig. (\ref{typ_cen_1}a).
(d) Spatial derivative of the density profile $s=d\ln n_{\rm n}/d\ln r$.
The initial cloud radius is $R_0 = 4.23$ pc.
A ``star'' on a curve, present only after a supercritical core forms, marks the 
instantaneous radius of the supercritical core. }
\end{figure*}

\subsection{Evolution of Central Quantities}

The model cloud would remain in the initial equilibrium state indefinitely if it were not for ambipolar diffusion. Figure (\ref{typ_cen_1}a) shows the evolution in time of the central number density  $n_{\rm n,c}$, normalized to its initial value, $n_{\rm n,c0} = 2711$ ${\rm cm^{-3}}$. The evolution of the cloud is followed through an enhancement of eleven orders of magnitude in central density, after which thermal ionization becomes important. Until the formation of a supercritical core,
denoted in the figure by the position of the ``star'' (at $n_{\rm n,c0} = 7.3 \times 10^{4}$ ${\rm cm^{-3}}$, $t=17$ Myr), the evolution is controlled by ambipolar diffusion and is relatively slow. The subsequent evolution is dynamic, although significantly slower than free-fall. During this phase, the central
density increases by more than ten orders of magnitude in less than
$1.5$ Myrs. 

Figure (\ref{typ_cen_1}b) exhibits the central abundances of all species
relative to the neutrals, $\chi_s \equiv n_s/n_{\rm n}$ (where
$s={\rm e, i, g^-, g^+, g^0}$), as functions of the central neutral density. At
low densities, corresponding to either the early stages of the
evolution of the central region or to the cloud envelope, the grains
are primarily negatively charged, because of the much more effective
electron (rather than ion) attachment onto the neutral
grains. This preferential depletion of electrons onto grains 
causes atomic and molecular ions to be slightly more
abundant than electrons. However, at these low densities, both
electrons and ions are much more abundant than the grain
species.  The fractional abundances of all grain species decrease
during the early, ambipolar-diffusion--controlled phase of the
evolution. This is the result of the attachment of charged grains to
magnetic field lines, which are ``left behind'' as the neutrals
contract through field lines under the action of their self-gravity. The same tendency is exhibited by the neutral grains because neutral and charged grains are well coupled through inelastic collisions. 

Above densities of $\sim 10^{10} {\rm \,\, cm^{-3}}$, the grains become
the main charge carriers. At these densities most grains are neutral, while
at higher densities positive grains become almost as abundant as
negative grains. Ions are considerably more abundant than
electrons, due to the faster depletion of electrons onto grains.

\begin{figure*}
\plotone{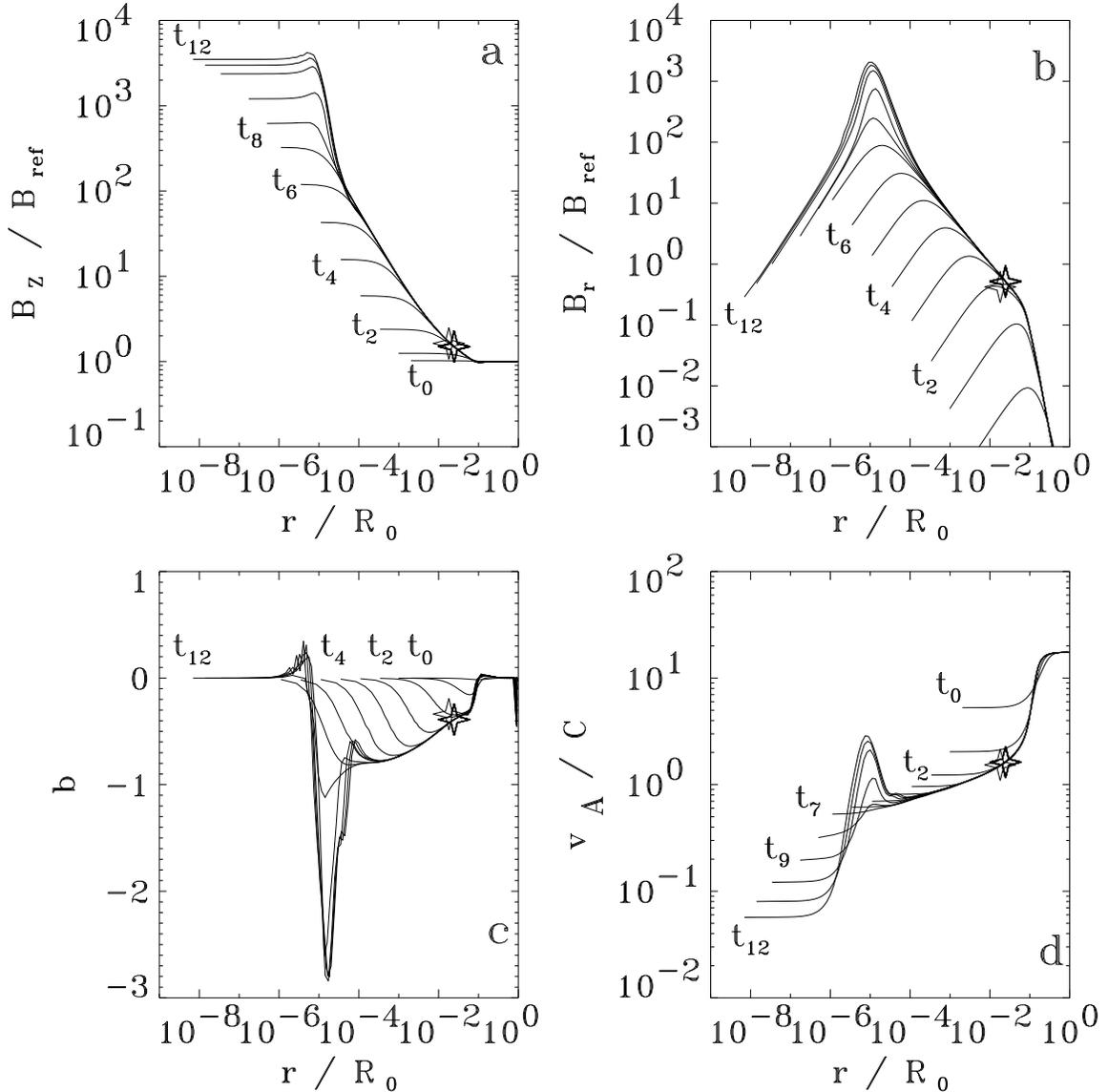}
\caption{\label{typ_rad_B} 
Radial profiles of physical quantities at different times, as in Fig. (\ref{typ_rad_rho}).
(a) $z-$component of the magnetic field  normalized to its value in
the reference state, $B_{\rm ref} = 35.26 {\rm \, \, \mu G}$. 
(b) $r-$component of the magnetic field normalized to $B_{\rm ref}$.
(c) Spatial derivative of the $z-$component of the magnetic field, $b
= d\ln B_{\rm z}/d\ln r$. 
(d) Alfv\'{e}n speed normalized to the sound speed of the reference state, 
$C = 0.188$ ${\rm km \, s^{-1}}$. 
A ``star'' on a curve, present only after a supercritical core forms, marks the 
instantaneous radius of the supercritical core. 
}
\end{figure*}

The abundances of all charged species decrease rapidly above a density
$\approx 3\times 10^{11} {\,\,\rm cm^{-3}}$, and then more slowly again above a
density $\approx 6 \times 10^{12} {\,\, \rm cm^{-3}}$. This behavior is the result of 
changes in the ionization rate, $\zeta$, shown in Figure (\ref{typ_cen_1}c).
At densities up to $\approx 6\times 10^{12} {\,\, \rm cm^{-3}}$, the
ionization is dominated by the contribution of cosmic rays. 
Its value is constant, $\zeta = 5\times 10^{-17} {\,\, \rm s^{-1}}$ for
densities up to $\approx  3\times 10^{11} {\,\,\rm cm^{-3}}$, when the
column density exceeds $\approx 100 {\,\, \rm g \, cm^{-2}}$
and the cosmic rays are appreciably attenuated. Hence, for higher
densities the ionization rate declines monotonically, until it is dominated by radioactive decays (mainly of $^{40}$K) above $n_{\rm n} \approx 6\times 10^{12} {\,\, \rm cm^{-3}}$, and reaches a constant value $\zeta = 6.9\times 10^{-23}
{\,\,\rm s^{-1}}$. Overall, between the initial constant value determined by cosmic rays and the high-density radioactive decay plateau, the ionization rate decreases
by almost 6 orders of magnitude. This decrease results in a very low
degree of ionization at these high-densities.

The central temperature as a function of central density is shown in Figure (\ref{typ_cen_1}d). For this model, the isothermal phase ($T = 10$ K) lasts up to
a density $n_{\rm opq} = 10^{11} {\,\,\rm cm^{-3}}$. For higher
densities, we assume an adiabatic equation of state with adiabatic
index equal to $5/3$ until a temperature of $200$ K is reached, which
occurs at a density  $9 \times 10^{12} {\,\, \rm cm^{-3}}$. At higher temperatures, the adiabatic index is changed to 7/5 to account for the excitation of the rotational degrees of freedom of molecular hydrogen.  A temperature of $10^3$ K is reached at a
central density of $5\times 10^{14} {\,\, \rm cm^{-3}}$, and the
calculation is stopped at this point, because thermal ionization (which is not accounted for) becomes important -- see also Footnote 1.

\begin{figure*}
\plotone{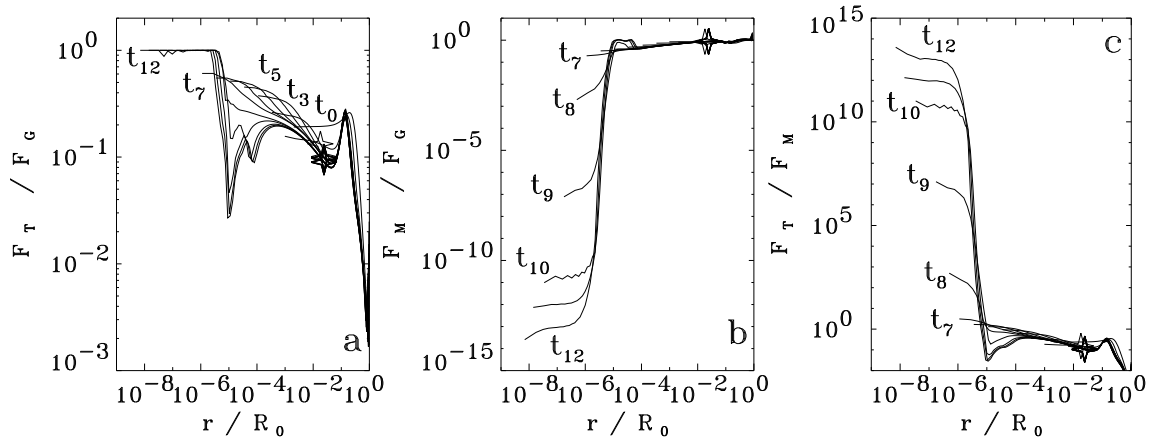}
\caption{\label{typ_rad_forces} 
Radial profiles of physical quantities at different times, as in Fig. (\ref{typ_rad_rho}).
(a) Ratio of thermal-pressure and gravitational forces. 
(b) Ratio of magnetic and gravitational forces.
(c) Ratio of thermal-pressure and magnetic forces.
A ``star'' on a curve, present only after a supercritical core forms, marks the 
instantaneous radius of the supercritical core. }
\end{figure*}

Figure (\ref{typ_cen_1}e) displays the central value of the
$z-$component of the magnetic field $B_{z, {\rm c}}$ normalized to its initial value,
$B_{z,{\rm c0}}=36$ $\mu G$, as a function of central density. During the
ambipolar-diffusion--controlled evolution (until the formation of a
supercritical core, marked by the ``star''), $B_{z,{\rm c}}$ increases only
by a factor of $1.6$. After that, a phase of near flux-trapping follows, during which $B_{z,{\rm c}}$ resembles a power-law $B_{z,{\rm c}} \propto n_{\rm n,c}^\kappa$ with $\kappa \lesssim 0.5$ (see Fig. \ref{typ_cen_1}f). At densities higher than a 
few $\times \, 10^{9} {\, \rm cm^{-3}}$,  $B_{z,{\rm c}}$ increases less rapidly with density, as the magnetic field starts to decouple from the matter
\citep{DM01}. This gradual process is abruptly interrupted because of
the change in the equation of state. However, the overall decrease in
$\kappa$ continues, and by the end of the run $\kappa$ has
decreased to a mean value of $0.1$. The oscillations exhibited by $\kappa$ after the onset of 
adiabaticity are the result of the system's attempt to achieve balance between thermal-pressure 
and gravitational forces in the hydrostatic core. 
For this reason, the central density and the central magnetic field exhibit small oscillations 
in time, which in turn manifest themselves as the ``noise'' in $\kappa$.\footnote{Note that the oscillations in $\kappa$ are physical and are not associated with the small-amplitude
post-shock oscillations appearing in the effective flux advection velocity $\bvec{v}_{\rm f}$ (see Fig. 
\ref{typ_rad_vb}). The latter are minimized through the use of artificial viscosity.}

\begin{figure*}
\plotone{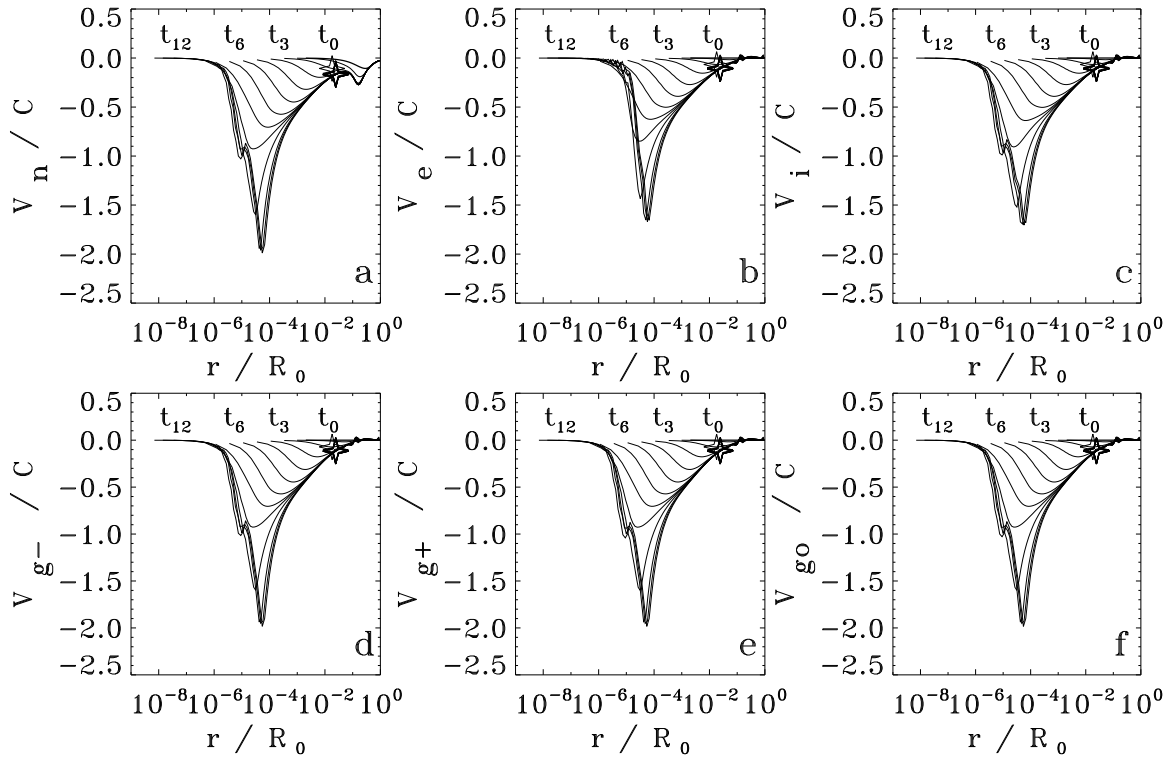}
\caption{\label{typ_rad_vel} 
Radial profiles of the velocities of the (a) neutrals; (b) electrons;
(c) ions; (d) negatively-charged grains; (e) positively-charged grains; and (f) neutral grains. All velocities are normalized to the sound speed of the
reference state ($C = 0.188$ ${\rm km s^{-1}}$). Different curves correspond to different times, as in Fig. (\ref{typ_rad_rho}).
A ``star'' on a curve, present only after a supercritical core forms, marks the 
instantaneous radius of the supercritical core. 
}
\end{figure*}

\subsection{Evolution of Spatial Structure of the Model Cloud}
\label{radresults}

Figures (\ref{typ_rad_rho}) - (\ref{typ_tadod}) show radial profiles of
physical quantities at different evolutionary times ($t_0 = 0$,  $t_1 = 14.37$, 
$t_2 = 17.9$, $t_3 = 18.35$, $t_4 = 18.42$, $t_5 = 18.432$, $t_6 = 18.434$, 
$t_7 = 18.435$, $t_8 = 18.4354$, $t_9 = 18.4358$, $t_{10} = 18.4362$,
$t_{11} = 18.4364$, $t_{12} = 18.4366$ - all times given in units of {\rm Myr}). 
Each curve corresponds to a time such that the central density has
increased by a factor of ten with respect to the previous curve; the
first curve corresponds to the equilibrium state and the last curve
to the time at which the temperature at the center of the model cloud
reaches $1000$ K.

\subsubsection{Densities and Temperature}
\label{dentemp}

Figure (\ref{typ_rad_rho}a) displays the column density of neutral
particles, in units of the central column density of the reference
state, as a function of radius at the thirteen different times given above. 
In the innermost part of the core, which is dominated by thermal-pressure forces that
smooth out all fluctuations, the column density is almost uniform. 
At late times, the column density outside the flat inner
region exhibits a profile resembling a broken power-law. It steepens 
appreciably  at small radii (but always outside the flat inner region)
where a hydrostatic protostellar core forms (see \S \, \ref{forces}).

Figure (\ref{typ_rad_rho}b) shows the radial dependence of the cloud
temperature (in ${\rm K}$). The temperature remains constant at $10
{\,\, \rm K}$ for all densities below $10^{11}$ ${\rm cm^{-3}}$.
At higher densities, it varies adiabatically with density, with an
adiabatic index equal to $5/3$ for $T<200{\, \rm K}$ and $7/5$ for
higher temperatures. The innermost uniform-density part of the core
also exhibits spatially uniform temperature, since $T \propto \rho^{\gamma -1}$.
Note that, because the core establishes a non-varying asymptotic density profile outside the pressure-dominated inner region, the radial
location of the onset of adiabaticity does not propagate outward in time. 

Radial profiles of the number density of neutral particles, normalized to the initial central number density ($n_{\rm n,c0} = 2711$ ${\rm cm^{-3}}$), are exhibited in Figure (\ref{typ_rad_rho}c). The logarithmic slope of the density, $s=d\ln n_{\rm n}/d\ln r$, is shown in Figure (\ref{typ_rad_rho}d). 
There are four prominent features exhibited by $s$. Inside the
supercritical core, $s$ approaches but does not become equal to $-2$
in the region $10^{-4} \lesssim r/R_0 \lesssim 10^{-2}$ (i.e., $10^{15} \, {\rm cm} \lesssim r \lesssim 10^{17} \, {\rm cm}$). Its mean value in this region is $s = -1.8$. In addition, there are three regions of significant steepening of the
density profile. The first is located just outside the supercritical
core ($r \sim 0.02 R_0$), and is a result of the loss of magnetic support for the
supercritical core, which contracts dynamically while the supported envelope
cannot replenish the inward moving mass fast enough to prevent the
steepening of the profile. The second and the third are located at
smaller radii. As will be discussed below (\S \, \ref{forces}), the second dip in 
$s$ (at $r \sim 2 \times 10^{-5} R_0$) corresponds to a balancing of the gravitational force by magnetic forces, while the third (at $r \sim 2 \times 10^{-6} R_0 $) 
corresponds to a balancing of the gravitational force by thermal-pressure forces.  
Note, however, that $s$ never becomes positive, and therefore the
density profile remains monotonic throughout the evolution. Figure (\ref{typ_rad_rho}c) by itself would give the impression that, outside the uniform-density central region, there are three different power-law regions: immediately beyond the central uniform density region, one might think that $s \approx -3$; further out, in the region $3 \times 10^{-5} R_0 \lesssim r \lesssim 10^{-2} R_0$, it appears that $s \approx -1.5$; and beyond the boundary of the supercritical core, $s \approx -2.5$. However, Figure (\ref{typ_rad_rho}d) reveals that the profile of the density is considerably more complicated than would be deduced by a visual inspection of Figure (\ref{typ_rad_rho}c).

\subsubsection{Magnetic Field and Alfv\'{e}n speed}
\label{Bfield}

The behavior of the magnetic field is demonstrated in Figure
(\ref{typ_rad_B}). The $z-$component of the magnetic field
is shown in Figure (\ref{typ_rad_B}a). At early times, and inside the supercritical core,  $B_z$ follows a pattern similar to that of the column and volume
densities,  with an innermost flat region followed by an almost power-law
region. The logarithmic slope varies slowly between $-0.5$ and $-1$ in the region
$10^{-4} \lesssim r/R_0 \lesssim 10^{-2}$ as is seen in Figure (\ref{typ_rad_B}c), which shows $b \equiv d\ln B_z / d\ln r$. Outside the supercritical core the profile
flattens significantly and $B_z$ is essentially uniform.

At later times, after the onset of adiabaticity in the inner part of the collapsing core, the spatial profile of $B_z$ steepens significantly. Outside the
location of the hydrostatic core, a local maximum appears in $B_z$ at $r \approx 10^{-5} R_0$, indicating a concentration of magnetic flux in this region. This behavior is demonstrated more dramatically in Figure (\ref{typ_rad_B}c), in which the logarithmic slope of $B_z$ acquires large negative values in the region where the $B_z$
profile steepens, while it crosses zero and becomes {\em positive} inward of the location of the local maximum.

Figure (\ref{typ_rad_B}b) shows the spatial dependence of the
$r-$component of the magnetic field, generated as a result of field-line deformation during the collapse. Inside the inner flat-density region, where the
magnetic field is almost uniform, $B_r$ declines rapidly to very small
values and it goes to zero as $r \rightarrow 0$. The rapid contraction outside the hydrostatic core induces a strong field-line deformation, and $B_r$ has a
sharp local maximum at $r \approx 10^{-5} R_0$, where its value becomes
comparable to, although still smaller than, that of $B_z$; 
$B_{r,{\rm max}}/B_z = 0.9$. In fact, $B_z > B_r$ at all radii and all times.

From Figures (\ref{typ_rad_B}a) and (\ref{typ_rad_B}c) it thus becomes clear that there are two distinct behaviors exhibited by the magnetic field inside and right outside the hydrostatic core. Inside the inner flat-density region, 
the field configuration becomes almost spatially uniform, while right outside 
this region a magnetic flux concentration is observed, with both the $z$ and $r$ 
components of the field exhibiting local maxima. The value of the $z-$component inside the hydrostatic core is $0.14 {\rm \, G}$, comparable to the protosolar magnetic field as derived from meteoritic data \citep{levy88,slp61,hr74}.

\begin{figure*}
\plotone{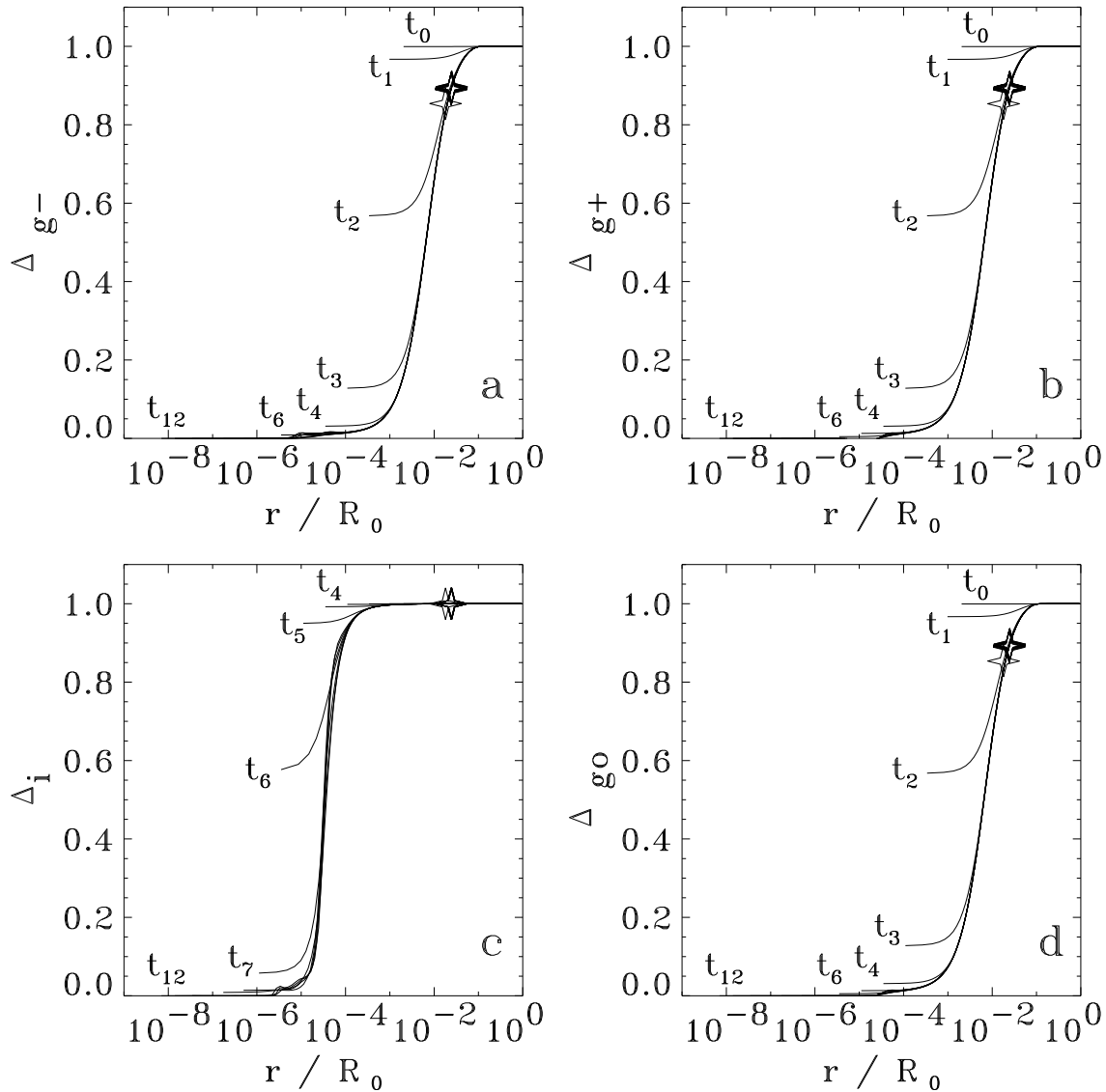}
\caption{\label{typ_rad_deltas} 
Radial profiles of attachment parameters for (a) the negatively-charged grains; 
(b) positively-charged grains; (c) ions; and
(d) neutral grains. Different curves correspond to different
times, as in Fig. (\ref{typ_rad_rho}).
A ``star'' on a curve, present only after a supercritical core forms, marks the 
instantaneous radius of the supercritical core. }
\end{figure*}

Figure (\ref{typ_rad_B}d) displays the Alfv\'{en} speed normalized to
the isothermal sound speed of the initial reference state. The local
maximum at the radius of the hydrostatic core is a result of the local
maximum in $B_z$; however the sound speed has also increased at
these radii as a result of the increased temperature. A most important
feature in the behavior of $v_A$ is that inside the supercritical
core its value becomes comparable to (or smaller than) that of the sound speed, 
which provides {\em an explanation for the thermalization of linewidths observed in molecular cloud cores} \citep{Baudry81,myb83}. In the magnetic star formation theory, 
the material motions responsible for the observed linewidths are attributed to
long-wavelength, standing Alfv\'{e}n waves (see Mouschovias 1987; Mouschovias \& Psaltis 1995; Mouschovias, Tassis, \& Kunz 2006), with a remarkable quantitative agreement between theory and observations.

\subsubsection{Forces}
\label{forces}

The radial dependence of the ratio of thermal-pressure and gravitational forces is shown in Figure (\ref{typ_rad_forces}a). The gravitational force dominates the thermal force everywhere (since the model cloud is thermally supercritical), except in 
the adiabatic core at late times, in which the thermal-pressure force exactly balances the gravitational force. The radius of the hydrostatic core is initially at $r = 5 \times 10^{-6} R_0$ and by the end of the run it has shrunk to $r = 2 \times 10^{-6} R_0$.

Figure (\ref{typ_rad_forces}b) shows the radial dependence of the
ratio of the magnetic and gravitational forces. Even inside the
dynamically contracting supercritical core, the magnetic force is an
appreciable fraction of the gravitational force. Hence {\em the dynamical contraction of the supercritical core is significantly slower than free-fall}. The magnetic force becomes much smaller than the gravitational force only inside the hydrostatic
core, where gravity is balanced by the thermal-pressure force. The magnetic force has a
local maximum with respect to the gravitational force at {\em the location
of the maximum of the magnetic field} seen in Figure (\ref{typ_rad_B}). This maximum
is located at $r \gtrsim 10^{-5} R_0 $, greater than the radius of the hydrostatic core,
defined as the radius inside which there is a balance between thermal-pressure and
gravitational forces (seen in Figure \ref{typ_rad_forces}a).

\begin{figure*}
\plotone{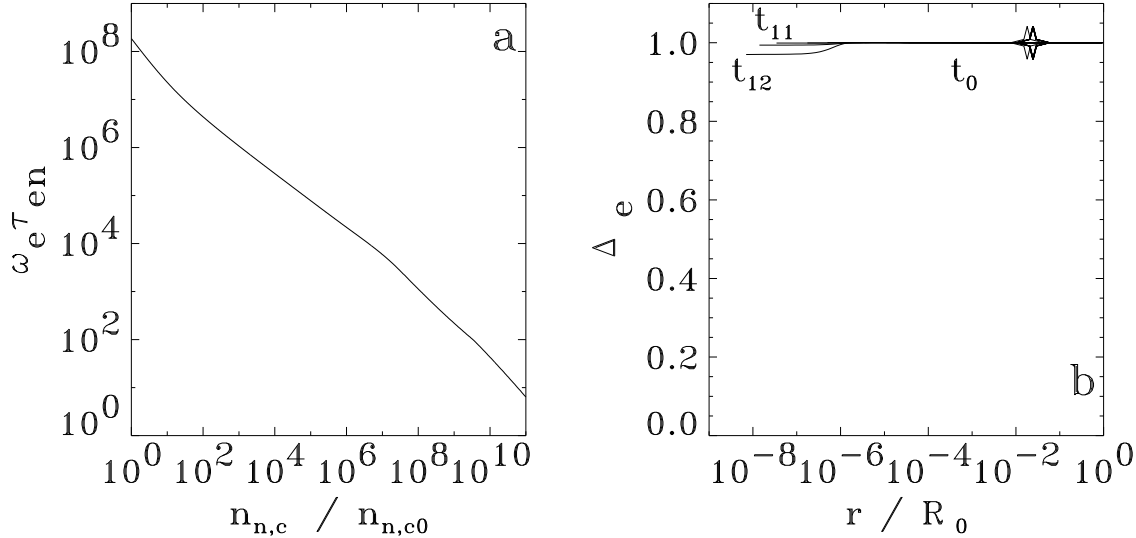}
\caption{\label{typ_rad_eattach} 
(a) Direct attachment parameter of the electrons at the center of the model cloud, 
as a function of the central neutral density enhancement. (b)
Radial profiles of the total electron attachment parameter, $\Delta_{\rm e}$.
Different curves correspond to different instants in
time, as in Fig. (\ref{typ_rad_rho}).
A ``star'' on a curve, present only after a supercritical core forms, marks the 
instantaneous radius of the supercritical core. }
\end{figure*}

Figure (\ref{typ_rad_forces}c) exhibits the spatial dependence of the ratio of
thermal-pressure and magnetic forces. The magnetic force is greater than
the thermal-pressure force (and hence provides the 
dominant opposition to gravity) everywhere except in the hydrostatic core, where
the thermal-pressure force dominates the magnetic force by many orders of
magnitude. Note, however, that just outside the hydrostatic core a local minimum 
in the thermal-to-magnetic force ratio forms, caused by the increased magnetic field 
strength shown in Figures (\ref{typ_rad_B}a) and (\ref{typ_rad_B}b).

\subsubsection{Velocities}
\label{velocities}

The velocity of the neutrals, normalized to the isothermal sound speed of the initial state, is plotted as a function of radius in Figure (\ref{typ_rad_vel}a). Before the onset of adiabaticity, the contraction speed is equal to a few tenths of the isothermal sound speed at the radius of the supercritical core and, at smaller radii, it increases to $0.5-1.0 \, \, C$. At yet smaller radii it decreases again and vanishes at $r=0$. After the innermost parts become adiabatic, the mass accumulating in the hydrostatic core causes an accelerated infall, and the magnitude of the velocity maximum becomes almost equal to $2 \, \,C$. The infall speed $v_{\rm n}$ decreases precipitously inward of the maximum (at $r/R_0 = 7 \times 10^{-5}$) and then again inward of $r = 9 \times 10^{-6} R_0$, at which a second, local maximum exists. This is the result of two shocks at the two locations, the outer one being magnetic in origin -- it coincides with the local maximum of the magnetic force -- and the inner one being an accretion shock at the boundary of the hydrostatic core.

Comparing the behavior of the velocities of the other species with
that of the neutrals, we observe that, at the late times of interest,
the velocity profiles of the ions and all grain species
(Figs. \ref{typ_rad_vel}c - \ref{typ_rad_vel}f) are almost identical with those
of the neutrals. Both the thermal and the magnetic shocks are experienced by the ions
and the grains. The situation is different for the electrons (Fig. \ref{typ_rad_vel}b), which
experience only the magnetic shock, while their velocity profile does not exhibit any significant feature at the location of the thermal shock. This is because the electrons are still well attached to the field lines (see discussion of Fig. \ref{typ_rad_eattach}b below) and are decelerated to almost zero velocity behind the magnetic shock. The margin for further electron deceleration due to the thermal shock (which occurs at smaller radii) is therefore much smaller than that for the other species. Hence, the electrons appear
to experience only the outermost, magnetic shock.

\subsubsection{Attachment Parameters}
\label{attach}

Figures (\ref{typ_rad_deltas}) and (\ref{typ_rad_eattach}) show radial
profiles of the attachment parameters for all species other than the neutrals at the thirteen different times. The attachment parameter for any species $s$ is defined as 
\beq
\Delta_s \equiv \frac{v_s - v_{\rm n}}{v_{\rm f} - v_{\rm n}}\,,
\eeq
where $v_{\rm f}$ is the effective flux advection velocity (the velocity at which the field lines appear to be moving inward). If $\Delta_s \approx 1$, then species $s$ is attached to the field lines. If $\Delta_s \ll 1$, then species $s$ is detached from the magnetic field lines and its motion follows that of the neutrals.

At low densities, all charged species and the neutral grains (which are well coupled to the charged grains through inelastic collisions) are attached to the field
lines. The grain species begin to detach first (at a radius $r/R_0 \approx 10^{-2} = 1.3 \times 10^{17} \, {\rm cm} = 8714$ AU, and a density of $n_{\rm n,c}\approx 10^{4} {\, \rm cm^{-3}}$), just as the magnetically (and thermally) supercritical core forms, and have completely detached by $n_{\rm n,c}\approx 10^{7} {\, \rm cm^{-3}}$. The ions begin to detach later, at a radius $r/R_{0}\approx 10^{-4} = 1.3 \times 10^{15} \, {\rm cm} \approx 87$ AU and density $n_{\rm n,c} \approx 10^{8} {\, \rm cm^{-3}}$, and have detached completely by $n_{\rm n,c} \approx 10^{10} {\, \rm cm ^{-3}}$. The ions are therefore still attached to the field lines at the location of the magnetic shock. 

\begin{figure*}
\plotone{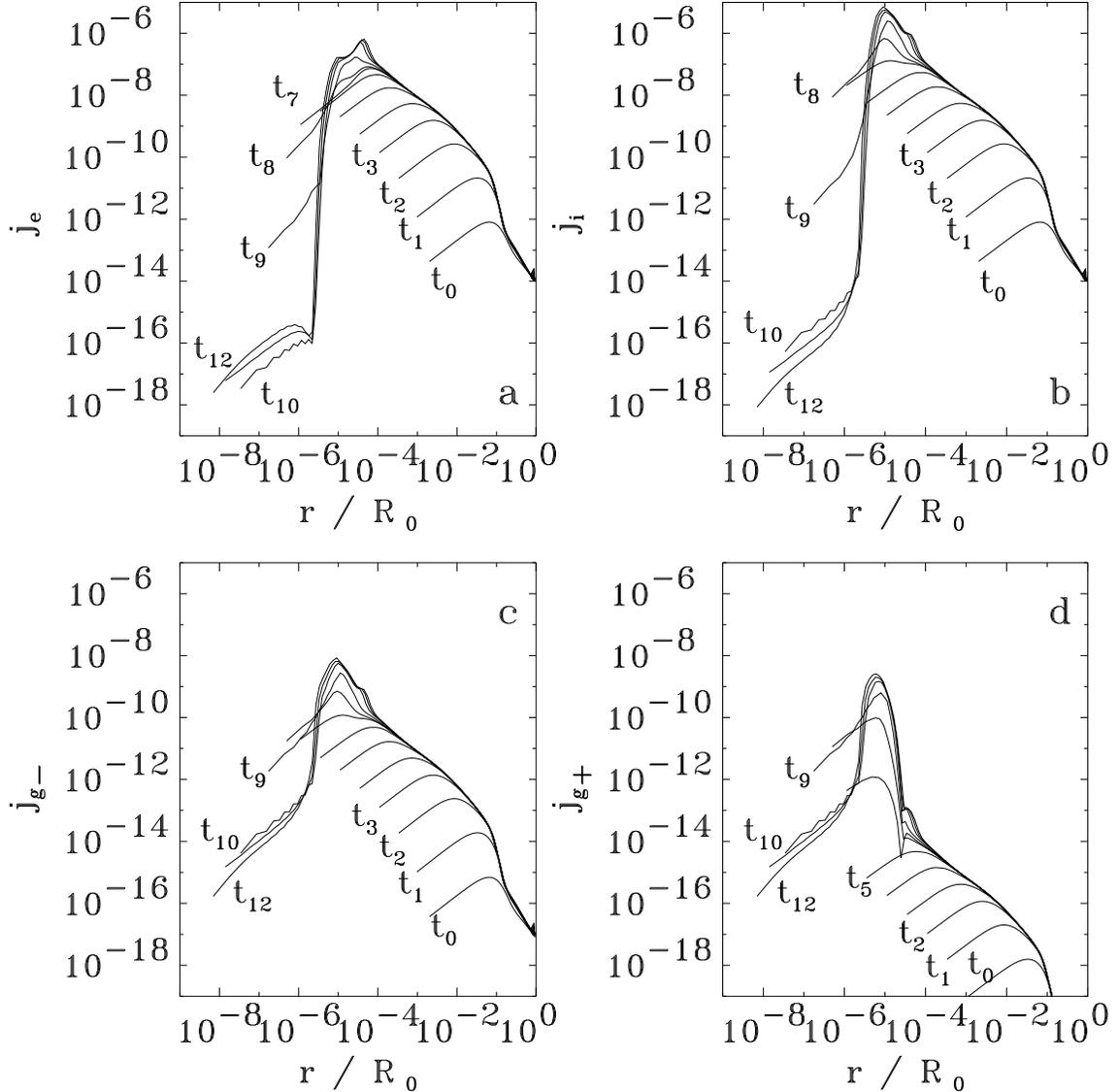}
\caption{\label{typ_rad_currents} 
Radial profiles of the electric current density in the $\phi-$direction 
carried by (a) the electrons; (b) the ions; (c) the negatively-charged grains; and
(d) the positively-charged grains. The units are
${\rm esu \,\, cm^{-2} \, \, s^{-1}}$ 
($1 \,\, {\rm esu \,\, cm^{-2} \, \, s^{-1}} = 3.3 \times 10^{-10} {\rm A \, \, cm^{-2}}$). 
Different curves correspond to different times, as in Fig. (\ref{typ_rad_rho}).
}
\end{figure*}

The electrons, as seen in Figure (\ref{typ_rad_eattach}b), have only partially detached from the field lines by the end of the run, when the temperature reaches 1000 K at the center of the cloud and thermal ionization is about to become important, increase the degree of ionization, and reattach the charged species to the field lines. Even at late times and in the innermost parts of the hydrostatic core, the electron attachment parameter does not
become smaller than $0.9$.  A similar conclusion can also be reached by examining the direct attachment parameter of the electrons, the value of which at the center of the cloud is plotted in Figure (\ref{typ_rad_eattach}a) as a function of the (normalized) central neutral density. It is the case that $\omega_{\rm e}\tau_{\rm en} \gtrsim 1$ essentially throughout the evolution.

\subsubsection{Contribution to Electric Current Density by Different Species}
\label{currents}

Figure (\ref{typ_rad_currents}) shows the contribution of different
charged species to the overall current, as a function of radius at different times. 
At early times, electrons and ions are the main current carriers, and
their contribution to the overall current density is comparable and approximately three orders of magnitude greater than that of positive and negative grains. At densities high enough so that the electron depletion onto the neutral grains causes the ion abundance to significantly exceed the abundance of the electrons, the ion current becomes greater than the
electron current by almost a factor of ten. Finally, in the innermost region, the current density decays sharply due to the increasing effect of Ohmic dissipation (see Fig. \ref{typ_tadod}). In this regime, grains become the main current carriers, their contribution
being almost two orders of magnitude greater than that of the electrons and ions.

\begin{figure*}
\plotone{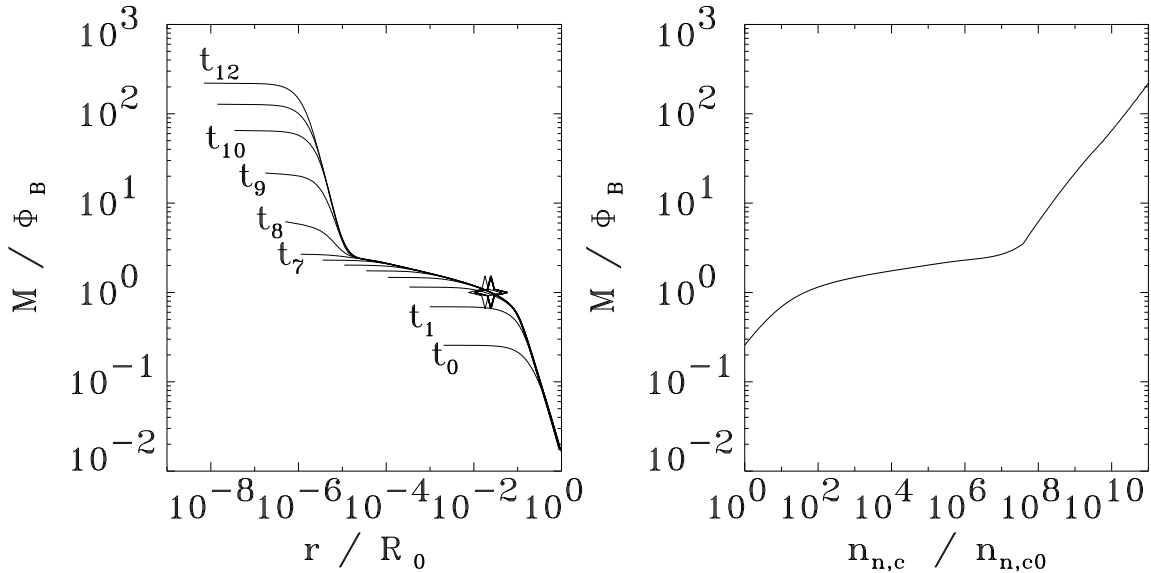}
\caption{\label{typ_m2f} 
(a) Radial profiles of the mass-to-flux ratio, normalized to the
  critical value for collapse, at different times (as in
  Fig. \ref{typ_rad_rho}).
(b) Central mass-to-flux ratio, normalized to the critical value, as a
  function of central density enhancement.
A ``star'' on a curve, present only after a supercritical core forms, marks the 
instantaneous radius of the supercritical core. }
\end{figure*}

\subsubsection{Mass-to-Flux Ratio and Drift Velocities}
\label{M2F}

The evolution of the mass-to-flux ratio in the model cloud is demonstrated in Figure (\ref{typ_m2f}).  Figure (\ref{typ_m2f}a) shows radial profiles of the mass-to-flux
ratio, normalized to the critical value for collapse, at the thirteen times given in \S \,  \ref{radresults}. The ``star'' marks the radius of the supercritical core, when the normalized mass-to-flux ratio becomes equal to unity. The mass-to-flux ratio is a three-slope power law with an inner flat region, corresponding to the flat-density, uniform-magnetic-field core. Inside the supercritical core but
outside the innermost flat region it exhibits two distinct behaviors: at outer radii it increases only mildly with decreasing radius, while at inner radii, inside the hydrostatic core, it increases sharply, and reaches a value a few hundred times greater than the critical value by the end of the run. Outside the supercritical core, the mass-to-flux ratio
decreases with radius and is independent of time.

A similar behavior can be seen in Figure (\ref{typ_m2f}b), where the normalized mass-to-flux ratio at the center of the cloud is plotted as a function of central density enhancement. Initially, during the ambipolar-diffusion--controlled phase, the central mass-to-flux ratio increases sharply, until it reaches the critical value, and a supercritical core forms. During the dynamical contraction stage, and while the ions are still attached to the magnetic field lines, the mass-to-flux ratio increases only mildly. Finally, after the ion detachment, a much steeper increase, by approximately two orders of magnitude, takes place.
This is an important contribution toward the resolution of the magnetic flux problem of
star formation, as the magnetic flux remaining at this stage approaches that observed in strongly magnetic stars. However, additional magnetic flux loss is required to account
for the entire observed range of magnetic fields in newborn stars. 

As discussed in detail in Paper I, the motion of the field lines can be described by an effective flux-advection velocity, $v_{\rm f}$, which is the apparent velocity {\em in the lab frame} with which the field lines are transported inward. When one or more charged species are
attached to the field lines, their velocity is $\approx v_{\rm f}$. When the field completely detaches from the matter, $v_{\rm f} \rightarrow 0$ and the field lines appear to be ``held in place'' as matter drifts inward through them.

{\it The magnetic field lines threading the supercritical core extend into the ``external'' medium
(the envelope) and the magnetic field is maintained by currents in the parent cloud envelope.} This is well justified because observations of the morphology of magnetic fields in and around molecular cloud ``clumps'' show fields of lengthscale greater than that of the clumps, threading the clumps and extending smoothly  into the parent cloud \citep{Hildeb99,mcg94}. {\rm Flux redistribution precesses (e.g., ambipolar diffusion) and dissipative processes (e.g., Ohmic dissipation) cannot therefore change the topology of any particular flux tube, or decrease the magnetic-field strength below that implied by the ``external'' currents.} They only limit the speed at which field lines are
transported inward, to a value smaller than that of the velocity of the neutrals, with the consequence that mass can accumulate in the central flux tubes, increasing
the mass-to-flux ratio there. Thus the {\em total} magnetic flux of the model cloud 
(but not that of the protostellar cores forming in it) is conserved.

\begin{figure*}
\plotone{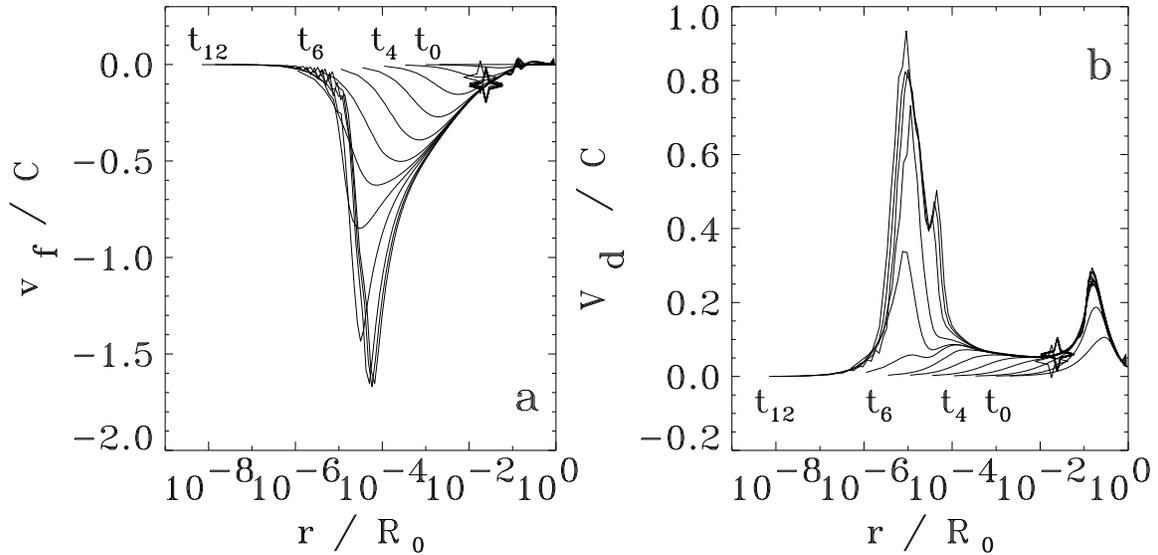}
\caption{\label{typ_rad_vb} 
Radial profile of (a) the effective flux-advection velocity, and (b) the drift velocity
(difference between the effective flux-advection velocity and the velocity of
the neutrals), in units of the sound speed of the reference state. 
Different curves correspond to different
times, as in Fig. (\ref{typ_rad_rho}).
A ``star'' on a curve, present only after a supercritical core forms, marks the 
instantaneous radius of the supercritical core. }
\end{figure*}

Radial profiles of $v_{\rm f}$ are shown in Figure (\ref{typ_rad_vb}a). Since, as we discussed above, the electrons do not detach from the magnetic field lines during the evolution, the radial and temporal behavior of $v_{\rm f}$ resembles that of $v_{\rm e}$ (see Fig. \ref{typ_rad_vel}b), with the prominent deceleration feature corresponding to the magnetic shock.

Figure (\ref{typ_rad_vb}b) shows the drift velocity between the field lines
and the neutrals, $v_{\rm f} - v_{\rm n}$. It exhibits a local
maximum in the model cloud envelope, where ambipolar diffusion continues to control 
the infall of the neutrals, with $|v_{\rm f}| \ll |v_{\rm n}|$ and 
$v_{\rm n} \approx -v_{\rm D}$. It then becomes small for a significant part of
the dynamically contracting supercritical core, where the field is
almost (although not exactly) frozen in the matter. Finally, it becomes
significant again in the inner parts of the core, where the degree of ionization is very low
and the collisional drag on the neutrals decreases.

\subsubsection{Mass and Magnetic Flux}

\begin{figure*}
\plotone{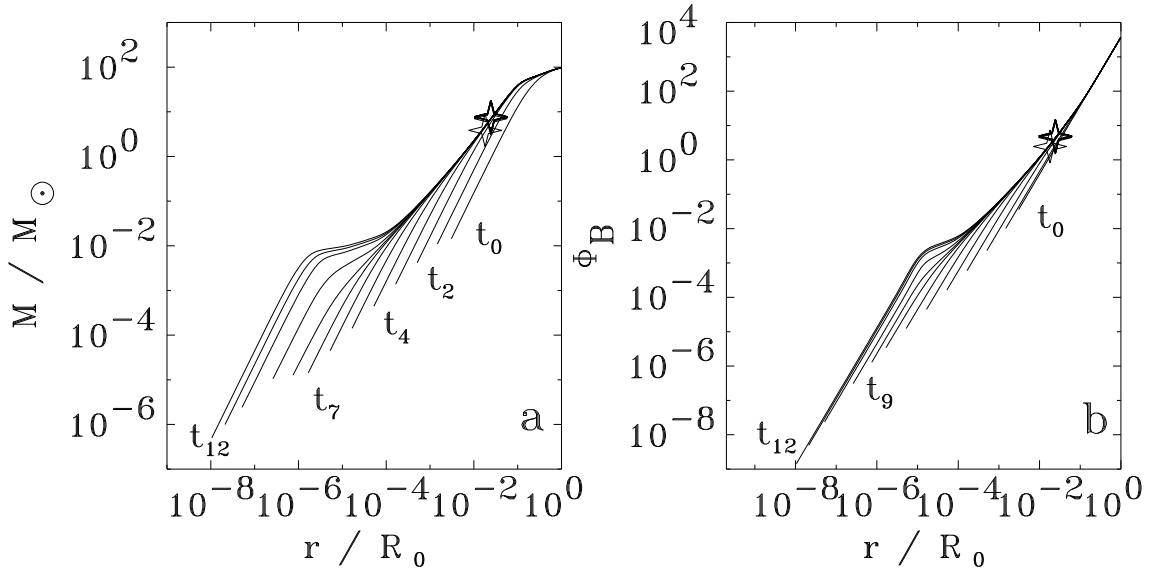}
\caption{\label{typ_rad_mf} 
Radial profiles of physical quantities at different 
times, as in Fig. (\ref{typ_rad_rho}). The distance from the center is normalized to the initial cloud radius, $R_0 = 4.23$ pc. Each line corresponds to a time such that the central density has increased by a factor of ten with respect to the previous curve. The
first line corresponds to the equilibrium state. The ``star'' on some curves marks the
location of the supercritical core.
(a) Cumulative mass.
(b) Cumulative flux. 
A ``star'' on a curve, present only after a supercritical core forms, marks the 
instantaneous radius of the supercritical core. }
\end{figure*}

Figures (\ref{typ_rad_mf}a) and (\ref{typ_rad_mf}b) show, respectively, 
radial profiles of the cumulative mass and magnetic flux at the thirteen 
different times given above. The ``star'' on a curve marks the location 
of the radius of the supercritical core, which contains only a few percent 
of the cloud mass. The mass included inside the {\em hydrostatic} core 
by the end of the run is $\approx 10^{-2} {\, \rm M_\odot}$.

At late times, both quantities exhibit a narrow plateau.
The change from an isothermal to an adiabatic equation of state at $n_{\rm n} \ge 10^{11} {\,\, \rm cm^{-3}}$ results in the formation of a high-density hydrostatic core, corresponding to a central mass concentration at the innermost part of the the supercritical core. The plateau in the mass profile appears because outside the hydrostatic core $r \approx 10^{-6} R_0$ the density of the disk declines rapidly (see Fig. \ref{typ_rad_rho}) and as a result the mass content of the plateau region ($10^{-6} R_0 \lesssim r \lesssim 2 \times 10^{-5} R_0$) is a very small fraction of the mass of the hydrostatic core.  

The deceleration of the infall of field lines at $r \approx 10^{-5} R_0$ (Fig. \ref{typ_rad_vb}a) results in a concentration of flux outside the hydrostatic core. This concentration dominates the magnetic flux profile in the plateau region $10^{-5} R_0 \lesssim r \lesssim 10^{-4} R_0$.

\subsubsection{Ambipolar Diffusion and Ohmic Dissipation}
\label{ADOD}

\begin{figure*}
\plotone{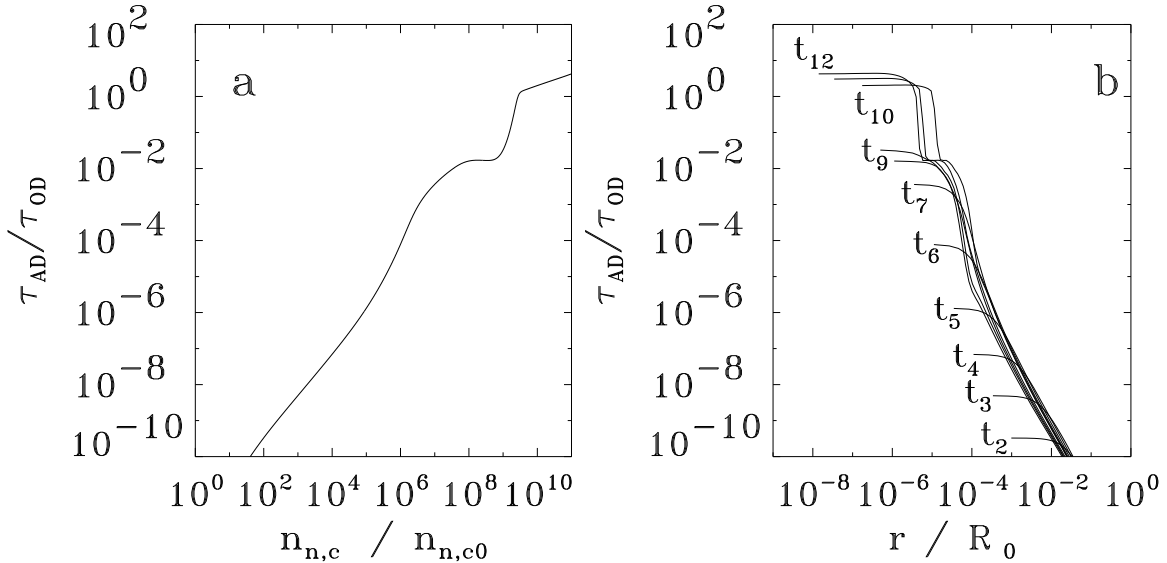}
\caption{\label{typ_tadod} 
(a) Central value of the ratio of ambipolar-diffusion and Ohmic-dissipation 
timescales as a function of central density. (b) Radial profile of the same 
ratio at different times, as in Fig. (\ref{typ_rad_rho}).
}
\end{figure*}

The relative effectiveness of ambipolar diffusion and Ohmic dissipation
in increasing the mass-to-flux ratio is demonstrated by Figure
(\ref{typ_tadod}), which shows the ratio of the ambipolar-diffusion and
Ohmic-dissipation timescales for the central cell as a function of
density (Fig. \ref{typ_tadod}a), and radial profiles of the same quantity at the thirteen different times (Fig. \ref{typ_tadod}b). At low densities and large radii, ambipolar diffusion dominates Ohmic dissipation. The two processes become equally important only at a density $5 \times 10^{12}$ ${\rm cm^{-3}}$, while at higher densities Ohmic dissipation operates on a timescale shorter than that of ambipolar diffusion. 

Interestingly, when the grains become the primary {\em charge} carriers, at
densities $\approx 2 \times 10^{10}$ ${\rm cm^{-3}}$, ambipolar diffusion still
dominates Ohmic dissipation as a flux-reduction mechanism, contrary 
to the expectations of \cite{nu86a,nu86b}. Instead, it is when the grains become
the main {\em current} carriers, at densities $\approx 10^{12}$ ${\rm cm^{-3}}$
that Ohmic dissipation dominates. The region in which this occurs coincides with the region in which the electric currents diminish ($r \lesssim 10^{-5} R_0$; see Fig. \ref{typ_rad_currents}).

At the location of the magnetic shock, ambipolar diffusion is always
more effective than Ohmic dissipation. This is consistent with the findings in 
\cite{TassisM05b}, where the magnetic shock was shown to be ambipolar-diffusion
driven, and the associated magnetic cycle was shown to repeat on the ambipolar-diffusion timescale.

\section{Conclusions and Discussion}\label{disc}

Decoupling of the magnetic field from the matter refers to the stage in which the  magnetic field becomes dynamically unimportant for the evolution of the neutral fluid and vice versa. We find that the magnetic forces begin to decline rapidly with respect to gravity, at a density $\approx 10^{11}$ ${\rm cm^{-3}}$ and 
radius $\lesssim 10^{-6} R_0 \approx 1 AU$. This marks the onset of magnetic decoupling. It begins while ambipolar 
diffusion is still more effective as a flux-reduction mechanism than Ohmic
dissipation. This is in agreement with the findings of \cite{DM01}; namely, {\it it is  ambipolar diffusion that initiates the decoupling of the magnetic field from the matter.} 

The field decoupling is not completed until all charged species detach from the magnetic field lines. This event does not occur in the model cloud since, by the end of this typical run, the electrons are still attached to the magnetic field lines. 
However, in Paper III we show that there are combinations of free parameters that lead to complete decoupling of the electrons at densities $\approx 10^{15}$ ${\rm cm^{-3}}$. In the present, representative calculation, this density is not achieved before the temperature reaches 1000 K. Still, the electrons are very few and they are very ineffective in transmitting the magnetic force to the neutrals through collisions. Therefore, the magnetic field in the innermost part of the contracting core is dynamically insignificant by the end of the run.

The grains begin to detach from the field lines at a central neutral density $n_{\rm n,c} \approx 10^{4}$ ${\rm cm^{-3}}$, at radius $\approx  10^{4}$ AU. Their detachment is complete by the time the central neutral density increases to $\approx 10^{8}$ ${\rm cm^{-3}}$. By contrast, the ions begin to detach when $n_{\rm n,c} \approx 10^{8}$ ${\rm cm^{-3}}$, in the region $r \lesssim  10^{4}$ AU, and their detachment is complete by the time $n_{\rm n,c} \approx 10^{11}$ ${\rm cm^{-3}}$. {\it The electrons in this representative model cloud, do not detach from the field lines by the time thermal ionization sets in.}

At densities $n_{\rm n,c} \lesssim 10^{9}$ ${\rm cm^{-3}}$ the dominant charge carriers are the ions and the electrons; they also are the dominant electric-current carriers, with the electrons being slightly less effective than the ions. In the protostellar fragment, 
at densities $n_{\rm n,c} > 10^{8}$ ${\rm cm^{-3}}$, the grains (negative and positive) 
are the dominant charge carriers. They also become the dominant current carriers above  
$n_{\rm n,c} \approx 10^{12}$ ${\rm cm^{-3}}$.

After the onset of the decoupling of the field from the matter, Ohmic dissipation ``catches up'' with ambipolar diffusion in effectiveness as a flux-reduction mechanism. This equality in the ambipolar-diffusion and Ohmic-dissipation timescales occurs at a mass-to-flux ratio $\approx 30$ times the critical value.

{\em The magnetic flux in the innermost part of the model cloud, even after Ohmic dissipation 
becomes dominant, is not destroyed}. The inward transport of the magnetic flux is halted and 
the magnetic field becomes spatially uniform. Indeed, this is exactly the behavior expected 
for a field anchored to the envelope of the cloud and supported by external currents, as is 
the case both in nature and in the
model cloud. From a mathematical point of view, once ambipolar diffusion and Ohmic dissipation 
erase spatial variations of the field in the innermost parts of the core, Ohmic dissipation is 
unable to operate further. This situation may change qualitatively after the magnetic field of 
the protostar becomes dipolar. Such an effect is not observed prior to the recoupling of the 
magnetic field due to thermal ionization; no significant pinching of the field lines occurs up 
to this point ($B_r \lesssim B_z$).

The magnetic field strength in the innermost part of the core at the end of the calculation is $0.14 
{\rm \, G}$, comparable to the magnetic field of the early solar system as measured in meteorites 
\citep{levy88,slp61,hr74}.

At the end of the calculation, the hydrostatic core at the center of the 
supercritical protostellar fragment contains about $0.01 \rm M_\odot$. The mass, of course, continues
to increase in time. If one wants to follow the evolution of the opaque core until its mass 
becomes comparable to a stellar mass, one would need to include new physics in the formulation of
the problem (e.g., radiative transfer, which would allow determination of the temperature evolution;
thermal ionization; and rotation -- see discussion below) as well as new chemistry. The evolution in 
this new regime will be addressed in a future publication. However, the evolution of the isothermal 
disk surrounding the opaque region until $1 {\, \rm M_\odot}$ has been accumulated at the center has been 
studied by Tassis \& Mouschovias (2005a,b). They found that accretion onto the forming protostar
occurs in a time-dependent, spasmodic fashion, due to a series of quasi-periodic magnetic shocks. In the 
work discussed here, we have seen the formation of the first of this series of shocks, which occurs 
outside the hydrostatic core, as a result of a {\em local} concentration of magnetic 
flux.\footnote{Note that the creation of the magnetic wall found by Tassis \& Mouschovias (2005b), which is responsible for the series of magnetically-driven shocks, is a local effect and does {\em not} depend on the amount of magnetic flux accumulated in the central region.} 
Our calculation demonstrated that, by the 
time thermal ionization becomes important, the mass-to-flux ratio of the hydrostatic protostellar core 
has increased by more than two orders of magnitude from its critical value. For comparison, 
observed protostellar magnetic fields imply mass-to-flux ratios two to four orders of magnitude 
above critical \citep{m87b} . Although the present result 
refers to a small protostellar mass, the work of Tassis \& Mouschovias (2005a,b) has demonstrated that
ambipolar diffusion alone in the accretion phase increases the mass-to-flux ratio of a $1 \, {\rm M_\odot}$ 
protostar by a similar amount. Interestingly, Shu et al. (2006) suggested that in the accretion phase, 
Ohmic dissipation alone cannot account for the required flux loss despite the fact that, in their 
semi-analytic approach, the
magnetic field has no dynamical effect on the system, and free-fall accretion is assumed. 

Inside the hydrostatic core, where the thermal-pressure force becomes the main opponent of the
gravitational force, the geometry tends to change from disklike to spherical (see, also, Desch \&
Mouschovias [2001], who studied the effect of this change on the results up to densities of $ 3 \times 
10^{12} {\rm cm}^{-3}$). The change in geometry changes the dependence of the gravitational field
from $r^{-3}$ to $r^{-2}$. However, in the mass and density regime we study, this effect is 
quantitatively small. Inside the hydrostatic core, the thermal-pressure force reduces the
contraction velocities to negligible values in any case. At the boundary of the hydrostatic core
at the end of the run, the difference between the gravitational field calculated under the disklike 
and spherical geometry assumptions is only $9\%$. Future calculations pursuing the detailed 
structure of the hydrostatic core at even higher densities will address this change in geometry. 

An associated issue is whether the assumption that centrifugal forces are negligible
is justified. As shown by Basu \& Mouschovias (1994), by the time a
supercritical core forms, enough angular momentum has been removed by magnetic braking
to render the centrifugal force insignificant relative to the gravitational force. After this time, 
as long as the disklike geometry is preserved, both gravity and the centrifugal force 
scale as $r^{-3}$, and their ratio is therefore constant. However, once the geometry changes from 
disklike to spherical, gravity scales as $r^{-2}$, and a centrifugal barrier can, in principle, be
present. In our case, the relevant question is whether such a barrier exists beyond the boundary of 
the hydrostatic core, where the inflow is still finite. Since the deviation of the gravitational 
force from its value calculated assuming disklike (instead of spherical) geometry is only $9\%$, 
no significant centrifugal forces are 
expected. This can be quantified easily. From Basu \& Mouschovias (1994)
Fig. 8, we find that the specific angular momentum corresponding to a central mass of 
$0.01 {\rm \, M}_\odot$ (the mass of our hydrostatic core at the end of the run) in their fiducial run 
is $10^{17} {\, \rm cm^{-2} s^{-1}}$. Under conservation of angular momentum,  
we can calculate the ratio of the centrifugal and gravitational forces at the boundary of the 
hydrostatic core assuming {\em spherical geometry}. We find this ratio to be $6 \times 10^{-4}$.
No realistic cloud rotation can give a significant centrifugal force at the stages of the evolution studied in the present investigation. Hence, neglecting the centrifugal force is justified throughout each of our simulations.

\acknowledgements{KT would like to thank Vasiliki Pavlidou 
for useful discussions and Glenn Ciolek for providing the base version of the code used 
in this work. The work of KT was supported in part by the University of Illinois 
through a Dissertation Completion Fellowship. This research was partially supported by a grant 
from the American Astronomical Society and NSF grants AST 02-06216 and AST 02-39759.}


\begin{thebibliography}

\bibitem [Basu \& Mouschovias (1994)]{BM94}
Basu, S. \& Mouschovias, T. Ch., 1994, 
ApJ, 432, 720
\bibitem[Baudry \etal (1981)]{Baudry81}
Baudry, A., Cernicharo, J., Perault, M., delaNoe, J., \& Despois, D. 1981,
  A\&A, 194, 101

\bibitem[Desch \& Mouschovias (2001)]{DM01}
Desch, S. J. \& Mouschovias, T. Ch. 2001, ApJ, 550, 314

\bibitem[Fiedler \& Mouschovias (1992)]{FM92}
Fiedler, R. A. \& Mouschovias, T. Ch. 1992, ApJ, 391, 199

\bibitem[Fiedler \& Mouschovias (1993)]{FM93}
---. 1993, ApJ, 415, 680

\bibitem[Herndon \& Rowe (1974)]{hr74}
Herndon, J. M. \& Rowe, M. W. 1974, Meteoritics, 9, 289

\bibitem[Hildebrand \etal (1999)]{Hildeb99}
Hildebrand, R. H., Dotson, J. L., Dowell, C. D., Schleuning, D. A., \&
  Vaillancourt, J. E. 1999, ApJ, 516, 834

\bibitem[Levy (1988)]{levy88}
Levy, E. H. 1988, in Meteorites and the Early Solar System, ed. J. F. Kerridge
  \& M. S. Matthews (Tucson: Univ. Arizona Press), 697

\bibitem[McGregor \etal (1994)]{mcg94}
McGregor, P. J., Harrison, T. E., Hough, J. H., \& Bailey, J. A. 1994, MNRAS,
  267, 755

\bibitem[Mouschovias (1987a)]{m87}
Mouschovias, T. Ch. 1987a, in Physical Processes in Interstellar Clouds, ed. G. E. Morfill \& M. Scholer (Dordrecht: Reidel), 453

\bibitem[Mouschovias (1987b)]{m87b}
Mouschovias, T. Ch. 1987b, in Physical Processes in Interstellar Clouds, ed.
  G. E. Morfill \& M. Scholer (Dordrecht: Reidel), 491
  
\bibitem[Mouschovias (1991)]{m91}
Mouschovias, T. Ch. 1991, ApJ, 373, 169

\bibitem[Mouschovias \& Paleologou (1986)]{mp86}
Mouschovias, T. Ch. \& Paleologou, E. V. 1986, ApJ, 308, 781

\bibitem[Mouschovias \& Psaltis (1995)]{mp95}
Mouschovias, T. Ch. \& Psaltis, D. 1995, ApJ, 444, L105

\bibitem[Mouschovias, Tassis, \& Kunz (2006)]{mtk06}
Mouschovias, T. Ch., Tassis K., \& Kunz, M. W. 2006, ApJ, 646, 1043

\bibitem[Myers \& Benson (1983)]{myb83}
Myers, P. C. \& Benson, P. J. 1983, ApJ, 266, 309

\bibitem[Nakano \& Umebayashi (1986a)]{nu86a}
Nakano, T. \& Umebayashi, T. 1986a, MNRAS, 218, 663

\bibitem[Nakano \& Umebayashi (1986b)]{nu86b}
---. 1986b, MNRAS, 221, 319

\bibitem[Stacey, Lovering, \& Parry (1961)]{slp61}
Stacey, F. D., Lovering, J. F., \& Parry, L. G. 1961, J. Geophys. Res., 66,
  1523

\bibitem[Shu et al. (2006)]{shu06}
Shu, F.H., Galli, D., Lizano, S., \& Cai, M. 2006, ApJ, 647, 382 

\bibitem[Tassis \& Mouschovias (2005a)]{TassisM05a}
Tassis, K. \& Mouschovias, T. Ch. 2005a, ApJ, 618, 769

\bibitem[Tassis \& Mouschovias (2005b)]{TassisM05b}
---. 2005b, ApJ, 618, 783

\bibitem[Tassis \& Mouschovias (2006)]{TassisM06a}
---. 2006 [paper I]

\bibitem[Tassis \& Mouschovias (2006)]{TassisM06c}
---. 2006 [paper III]

\bibitem[Toomre(1963)]{Toomre63}
Toomre, A. 1963, ApJ, 138, 385

\end{thebibliography}
\end{document}